\documentclass[12pt]{iopart}
\usepackage{iopams}
\usepackage{graphicx}
\usepackage{caption}
\usepackage{subcaption}
\usepackage{hyperref}
\usepackage{hypernat}
\usepackage{color}

\begin{document}
\title[Discrete-time random walks and L\'evy flights on arbitrary networks]{Discrete-time random walks and L\'evy flights on arbitrary networks: when resetting becomes advantageous?}
\author{Alejandro P. Riascos${}^{1}$, Denis Boyer${}^{1}$ and Jos\'e L. Mateos${}^{1,2}$}  
\address{${}^1$Instituto de F\'isica, Universidad Nacional Aut\'onoma de M\'exico, 
	C.P. 04510, Ciudad de M\'exico, M\'exico\\
	${}^2$ Centro de Ciencias de la Complejidad, Universidad Nacional Aut\'onoma de M\'exico, C.P. 04510, Ciudad de M\'exico, M\'exico}

\begin{abstract}
The spectral theory of random walks on networks of arbitrary topology can be readily extended to study random walks and L\'evy flights subject to resetting on these structures. When a discrete-time process is stochastically brought back from time to time to its starting node, the mean search time needed to reach another node of the network may be significantly decreased. In other cases, however, resetting is detrimental to search. Using the eigenvalues and eigenvectors of the transition matrix defining the process without resetting, we derive a general criterion for finite networks that establishes when there exists a non-zero resetting probability that minimizes the mean first passage time at a target node. Right at optimality, the coefficient of variation of the first passage time is not unity, unlike in continuous time processes with instantaneous resetting, but above 1 and depends on the minimal mean first passage time. The approach is general and applicable to the study of different discrete-time ergodic Markov processes such as L\'evy flights, where the long-range dynamics is introduced in terms of the fractional Laplacian of the graph. We apply these results to the study of optimal transport on rings and Cayley trees.
\end{abstract}

\section{Introduction}
Random walks are ubiquitous in nature and find applications to a broad range of fields. The decomposition of a stochastic trajectory into a succession of discrete steps naturally describes many processes like diffusion, chemical reactions, animal movements, human mobility, and search processes in general. Furthermore, it is often convenient to represent the space on which a dynamical process takes place as a network (see \cite{ReviewJCN_2021} for a recent review and references therein). Random walks that transit between nodes are relevant to many problems and constitute the natural framework to study diffusive transport in regular and irregular structures \cite{VespiBook,Hughes,Lovasz1996}. Complex network exploration by random walks can be defined through hops to nearest neighbors \cite{ReviewJCN_2021,NohRieger2004,Tejedor2009PRE,MasudaPhysRep2017} or with long-range jumps between distant nodes \cite{ReviewJCN_2021,RiascosMateos2012,RiascosMateosFD2014,FractionalBook2019}. The understanding of the relation between the random walk dynamics and the network topology requires a particular treatment in terms of matrices and spectral methods \cite{NewmanBook,GodsilBook, VanMieghem2011}.
\\[2mm]
In part due to their relevance for random searches, there has been in recent years a marked interest for processes with resetting or restart. Processes under resetting are related to a variety of phenomena such as animal foraging \cite{luca,Pal_PhysRevResearch_2020}, genome conversion \cite{lev}, extinctions in population dynamics \cite{extinction}, problem solving in computer science \cite{montanari}, data search \cite{brin1998,leskovec2014,ermann2015} or catalytic reactions \cite{reuveni2014role,landau} (see \cite{EvansReview2019} for a review on resetting processes). Bringing a given process back to its starting point at regular or random time intervals profoundly affects several important static and dynamic observables. Notably, the average time needed to reach a fixed target state for the first time can be often minimized with respect to the resetting rate \cite{evans2011diffusion,Evans2011JPhysA,reuveni2016optimal}. The theory of resetting processes has witnessed significant advances in the last decade \cite{EvansReview2019}. Most of the works in the field have focused on continuous-time resetting in simple geometries, while  less attention has been given to discrete-time processes. One may emphasize the understanding brought by basic examples such as the one-dimensional Brownian motion under stochastic \cite{evans2011diffusion,Evans2011JPhysA,reuveni2016optimal} or periodic \cite{pal2016diffusion} resetting, its extensions to more general resetting protocols \cite{nagar2016diffusion,Bhat2016JStat,chechkin2018random,eule} or to anomalous diffusive systems \cite{kusmierz2015optimal,kusmierz2019subdiffusive,maso2019transport}. 
Although they remain relatively few, some recent experimental works using micro-sized particles have not only confirmed several theoretical predictions  \cite{ciliberto,israel} but also motivated further research on new phenomena associated with resetting when physical constraints are present \cite{ciliberto,mercado,trap}. 
\\[2mm]
Studies on Brownian motion in bounded one-dimensional intervals \cite{christou,Pal_PRE2019_interval}, in spherical domains in higher dimensions \cite{chen}, or on the infinite line in the presence of a drift \cite{montero2013monotonic,ray2019peclet}, have shown that resetting does not always expedite the completion of a search process, causing a significant slowdown instead. In the context of continuous time processes, a surprisingly simple and general criterion tells whether the mean search time of a process can be decreased by stochastic resetting or not \cite{reuveni2016optimal,chechkin2018random,PalFP_PRL_2017,crit3}: resetting is beneficial only if the fluctuations of the search time in the absence of resetting are sufficiently large. More precisely, the coefficient of variation of the first passage time in the resetting-free process must be above unity. As a corollary, if a non-zero optimal resetting rate exists, the coefficient of variation is unity at optimality. These results follow from the renewal structure of restart processes, where the memory of past excursions is erased after each restart, which is also assumed to be instantaneous. The purpose of this work is to extend these ideas by using a different methodology which is well adapted to study discrete-time processes taking place on arbitrary networks.
\\[2mm]
The study of discrete-time processes under the influence of resetting is relatively recent. Several results have been derived in those cases, mostly on the line or one-dimensional lattices \cite{Kusmierz2014PRL, christophorov2020peculiarities,Avrachenkov2014,Avrachenkov2018,Touchette_PRE2018,MajumdarSabhapandit_PRE2015,Montero_PRE_2016,villarroel2021semi, ResetNetworks_PRE2020,Wald_PRE2021,bonomo2021first,bonomo2021polya}. Reference \cite{Kusmierz2014PRL} provides a general formula that relates the survival probability of a generic discrete-time process subject to geometric resetting to the one without resetting. It also provides a formula for the first passage time between two nodes. Reference \cite{bonomo2021first} provides analog relations to the ones given in \cite{Kusmierz2014PRL} for a generic process under a generic resetting distribution. 
In a recent work, we have considered discrete-time random walks subject to resetting on arbitrary connected networks and established general relationships between several basic quantities and the spectral representation of the transition matrix that defines the random walk without resetting \cite{ResetNetworks_PRE2020}. It is worth mentioning that the connection between a resetting problem and the underlying resetting-free process is a common property under many different resetting protocols,  both in continuous and discrete-time (see, {\it e.g.}, \cite{Pal_PhysRevResearch_2020,EvansReview2019,Kusmierz2014PRL,evans2018run}). The results in Ref. \cite{ResetNetworks_PRE2020}  were further extended to the case of resetting to multiple nodes \cite{MultipleResetPRE_2021}. 
\\[2mm]
In this research, we analyze the optimal resetting of discrete-time random walks on networks. In Section \ref{RW_reset1}, we present general definitions and recall previous results on these processes. In particular, the eigenvalues and eigenvectors of the transition matrix that describes a random walk with resetting can be expressed in terms of the same quantities for the process without resetting. This section summarizes the methods introduced in Refs. \cite{Touchette_PRE2018,ResetNetworks_PRE2020} and the analytical expressions for the stationary distribution and the mean first passage time (MFPT). In Section \ref{Sec_generalresults} we present the main contribution of our research, were we derive a condition for optimal resetting (such that the mean first passage time to a given node is minimum) and using the second moment of the first passage time distribution, we deduce a condition that must be fulfilled at optimality. We also show that the introduction of resetting improves the mean search time when an inequality for the first passage time fluctuations of the resetting-free process is fulfilled. The spectral methods introduced for the analysis of random walks with resetting are general and can be applied to other discrete-time ergodic Markovian processes. In Section \ref{Sec_examples} we illustrate our general theory 
from which we obtain numerical values for the optimal resetting probability  in the case of local random walks on rings and Cayley trees. We further extend the results to L\'evy flights on rings, these processes being generated by the fractional Laplacian of the graph. We present the conclusions in Section \ref{sec:conclusions}.


\section{Random walks with resetting}
\label{RW_reset1}
We consider a random walker on an arbitrary connected network with $N$ nodes $i=1,\ldots ,N$. By definition, time is discrete, {\it i.e.}, $t=0,1,2,\ldots$. The walker starts at $t=0$ from the node $i$ and chooses between two types of rules at each time step: with probability $1-\gamma$, the walker randomly jumps from the node currently occupied to a different node of the network, or, with the complementary probability $\gamma$, resets to a fixed node $r$. In the first case, or for the dynamics without resetting ($\gamma=0$), the probability to hop to $m$ from $l$ is given by $w_{l\to m}$ or $w_{l,m}$, and we assume that the random walk is ergodic. We introduce the transition matrix $\mathbf{W}$, whose elements $w_{l,m}$ represent the probabilities to jump from node $l$ to node $m$ ($l,m=1,\ldots,N$). The matrix $\mathbf{W}$ can include transitions between connected nodes only, or long-range jumps between distant nodes \cite{ReviewJCN_2021}. The matrix
$\mathbf{\Pi}(r;\gamma)$ with elements $\pi_{l \to m}(r;\gamma)\equiv (1-\gamma) w_{l\to m}+\gamma\,\delta_{rm}$ ($\delta_{rj}$ is the Kronecker delta), completely describes the process subject to resetting, which is able to reach all the nodes of the network if the resetting probability $\gamma$ is $<1$.
\\[2mm]
The occupation probability of the process defined above follows the master equation
\begin{equation}\label{mastermarkov}
P_{ij}(t+1;r,\gamma) = \sum_{l=1}^N  P_{il} (t;r,\gamma) \pi_{l\to j}(r;\gamma).
\end{equation}
Here $P_{ij}(t;r,\gamma)$ denotes the probability to find the walker on node $j$ at time $t$, considering the initial position $i$ at $t=0$, the resetting node $r$ and the resetting probability $\gamma$.
\\[2mm] 
In Dirac’s notation, the formal solution of Eq. (\ref{mastermarkov}) that fulfils the initial condition $\delta_{ij}$ is 
$P_{ij}(t;r,\gamma)=\langle i|\mathbf{\Pi}(r;\gamma)^t|j\rangle$, where $|j\rangle$ denotes the $N$-component vector that has all its entries equal to 0 except 
the $j$-th one, which is equal to 1 ($\langle i|$ is the transpose of $|i\rangle$).
The matrices $\mathbf{W}$ and $\mathbf{\Pi}(r;\gamma)$ are stochastic matrices (their elements $a_{l, m}$ are such that $\sum_{m=1}^N a_{l,m}=1$), not necessarily symmetric, and the knowledge of their eigenvalues and eigenvectors allows the calculation of several important quantities, such as the occupation probability at all times $P_{ij}(t;r,\gamma)$, the stationary distribution at $t=\infty$, as well as the MFPT to any node. 
\\[2mm]
Let us denote the eigenvalues of  $\mathbf{W}$ as $\lambda_l$ (where $\lambda_1=1$), and its right and left eigenvectors as $\left|\phi_l\right\rangle$ and $\left\langle\bar{\phi}_l\right|$, respectively, for $l=1,2,\ldots,N$. The eigenvectors satisfy the conditions $\left\langle\bar{\phi}_l|\phi_m\right\rangle=\delta_{lm}$ and
$\sum_{l=1}^N \left|\phi_l\right\rangle \left\langle\bar{\phi}_l\right|=\mathbb{I}$, where 
$\mathbb{I}$ is the $N\times N$ identity matrix. These spectral properties can be either obtained numerically or analytically in some cases. On the other hand, the eigenvalues of $\mathbf{\Pi}(r;\gamma)$ are denoted as $\zeta_l(r;\gamma)$ and its right/left eigenvectors as $\left|\psi_l(r;\gamma)\right\rangle$ and $\left\langle\bar{\psi}_l(r;\gamma)\right|$. The eigenvalues $\zeta_l(r;\gamma)$ for the dynamics with restart are related with $\lambda_l$ through the expression (see Ref. \cite{ResetNetworks_PRE2020} for a detailed discussion):
\begin{equation}\label{eigvals_zeta}
\zeta_l(r;\gamma)=
\left\{
\begin{array}{ll}
1 \qquad &\mathrm{for}\qquad l=1,\\
(1-\gamma)\lambda_l \qquad &\mathrm{for}\qquad l=2,3,\ldots, N.
\end{array}\right.
\end{equation}
Furthermore, the left eigenvectors of $\mathbf{\Pi}(r;\gamma)$ are given by  \cite{ResetNetworks_PRE2020}
\begin{equation}\label{psil1}
\left\langle\bar{\psi}_1(r;\gamma)\right|=\left\langle\bar{\phi}_1\right|
+\sum_{m=2}^N\frac{\gamma}{1-(1-\gamma)\lambda_m}\frac{\left\langle r|\phi_m\right\rangle}{\left\langle r|\phi_1\right\rangle}\left\langle\bar{\phi}_m\right|.
\end{equation}
Recall that $|r\rangle$ denotes the vector that has all its components equal to 0 except 
the $r$-th one, which is equal to 1. For $l=2,\ldots,N$, one gets
$\left\langle\bar{\psi}_l(r;\gamma)\right|=\left\langle\bar{\phi}_l\right|$. The right eigenvectors of $\mathbf{\Pi}(r;\gamma)$ are given by
$\left|\psi_1(r;\gamma)\right\rangle=\left|\phi_1\right\rangle$ and
\begin{equation}\label{psirl_reset}
\left|\psi_l(r;\gamma)\right\rangle=
\left|\phi_l\right\rangle-\frac{\gamma}{1-(1-\gamma)\lambda_l}\frac{\left\langle r|\phi_l\right\rangle }{\left\langle r|\phi_1\right\rangle}\left|\phi_1\right\rangle,
\end{equation}
for $l=2,\ldots,N$. In this way, assuming that the eigenvalues and eigenvectors for the dynamics without resetting $\mathbf{W}$ are known analytically or numerically,  we have \cite{ResetNetworks_PRE2020}
\begin{equation}
P_{ij}(t;r,\gamma)=P_j^\infty(r;\gamma)+\sum_{l=2}^N(1-\gamma)^t\lambda_l^t\left[\left\langle i|\phi_l\right\rangle \left\langle\bar{\phi}_l|j\right\rangle-\gamma\frac{\left\langle r|\phi_l\right\rangle \left\langle\bar{\phi}_l|j\right\rangle}{1-(1-\gamma)\lambda_l} \right]. \label{Pijspect}
\end{equation}
The first term of the r.h.s. in Eq. (\ref{Pijspect}) defines the long time stationary distribution, $P_j^\infty(r;\gamma)=\left\langle i\left|\psi_1(r;\gamma)\right\rangle \left\langle\bar{\psi}_1(r;\gamma)\right|j\right\rangle$, which is the only contribution that does not decay to zero as $t\rightarrow\infty$.
From these expressions, 
the occupation probability $P_{ij}(t;r,\gamma)$ at times $t=1,2,\ldots$, can thus be expressed in terms of the eigenvalues and eigenvectors of $\mathbf{W}$. In particular, in the long time limit, one obtains the non-equilibrium steady-state distribution 
\begin{equation}\label{Pinfvectors_1}
P_j^\infty(r;\gamma)=P_j^\infty(0)+\gamma\sum_{l=2}^N\frac{\left\langle r|\phi_l\right\rangle \left\langle\bar{\phi}_l|j\right\rangle}{1-(1-\gamma)\lambda_l}.
\end{equation}
The term $P_i^\infty(0)\equiv\left\langle i|\phi_1\right\rangle \left\langle\bar{\phi}_1|j\right\rangle$ denotes the stationary probability distribution of the random walk without resetting. In the case of unbiased nearest-neighbour random walks (see Section \ref{Normal_Rings}), it is given by the general expression $P_j^\infty(0)=k_j/\sum_{l=1}^N k_l$, where $k_j$ is the number of neighbours of $j$
\cite{NohRieger2004,MasudaPhysRep2017}.
\\[2mm]
The mean first passage time of the random walker at node $j$, given a starting node $i$, resetting node $r$ and resetting probability $\gamma$, is denoted as $\langle T_{ij}(r;\gamma)\rangle$. Its expression is deduced in \cite{ResetNetworks_PRE2020} and given by
\begin{equation}\label{MFPT_resetSM}
\left\langle T_{ij}(r;\gamma)\right\rangle=\frac{\delta_{ij}}{P_j^\infty(r;\gamma)}\\+
\frac{1}{P_j^\infty(r;\gamma)}\sum_{l=2}^N\frac{
\left\langle j|\phi_l\right\rangle \left\langle\bar{\phi}_l|j\right\rangle-\left\langle i|\phi_l\right\rangle \left\langle\bar{\phi}_l|j\right\rangle
}{1-(1-\gamma)\lambda_l}.
\end{equation}
\section{Optimal resetting for random walks on networks}
\label{Sec_generalresults}
\subsection{Optimal resetting}
\label{SecOptimalReset}
In this section, we consider random walks with restart to the initial node, or $r=i$, and seek the optimal resetting probability $\gamma^\star$ that minimizes the MFPT given by
\begin{equation}\label{MFPTresetorigin}
\left\langle T_{ij}(\gamma)\right\rangle\equiv \left\langle T_{ij}(i;\gamma)\right\rangle=
\frac{1}{P_j^\infty(i;\gamma)}\sum_{l=2}^N\frac{
\left\langle j|\phi_l\right\rangle \left\langle\bar{\phi}_l|j\right\rangle-\left\langle i|\phi_l\right\rangle \left\langle\bar{\phi}_l|j\right\rangle
}{1-(1-\gamma)\lambda_l},
\end{equation}
where we have assumed $i\neq j$, the case of interest in the following.
In a more compact notation, motivated by the form of the stationary distribution $P_j^\infty(r;\gamma)$ in Eq. (\ref{Pinfvectors_1}), one may represent Eq. (\ref{MFPTresetorigin}) as
\begin{equation}\label{MFPTCS}
\left\langle T_{ij}(\gamma)\right\rangle=
\frac{\mathcal{C}_{ij}(\gamma)}{P_j^\infty(0)+\gamma\, \mathcal{S}_{ij}(\gamma)},\qquad i\neq j,
\end{equation}
where
\begin{equation}\label{CoeffCij}    
\mathcal{C}_{ij}(\gamma)\equiv \sum_{l=2}^N\frac{1
}{1-(1-\gamma)\lambda_l}\left[
\left\langle j|\phi_l\right\rangle \left\langle\bar{\phi}_l|j\right\rangle-\left\langle i|\phi_l\right\rangle \left\langle\bar{\phi}_l|j\right\rangle\right],
\end{equation}
and
\begin{equation}
\label{CoeffSij}
\mathcal{S}_{ij}(\gamma)\equiv \sum_{l=2}^N\frac{1}{1-(1-\gamma)\lambda_l}\left\langle i|\phi_l\right\rangle \left\langle\bar{\phi}_l|j\right\rangle.
\end{equation}
Under this form, by imposing $\frac{d}{d\gamma}    \left\langle T_{ij}(\gamma)\right\rangle\Big|_{\gamma=\gamma^\star}=0$, we obtain a relation satisfied by the optimal resetting probability $\gamma^\star$ (if it is non-zero):
\begin{equation}\label{Opt_Cond1}
\mathcal{C}'_{ij}(\gamma^\star)[P_j^\infty(0)+\gamma^\star\, \mathcal{S}_{ij}(\gamma^\star)]-\mathcal{C}_{ij}(\gamma^\star)[\gamma^\star\, \mathcal{S}^\prime_{ij}(\gamma^\star)+\mathcal{S}_{ij}(\gamma^\star)]=0,
\end{equation}
where $\mathcal{F}'(\gamma^\star)$ denotes $\frac{d}{d\gamma}\mathcal{F}(\gamma)\Big|_{\gamma=\gamma^\star}$.  The derivatives $\mathcal{C}'_{ij}(\gamma)$ and $\mathcal{S}'_{ij}(\gamma)$ are given by
\begin{equation}\label{CoeffCijprime}
\mathcal{C}'_{ij}(\gamma)=-\sum_{l=2}^N\frac{\lambda_l}{[1-(1-\gamma)\lambda_l]^2}\left[
\left\langle j|\phi_l\right\rangle \left\langle\bar{\phi}_l|j\right\rangle-\left\langle i|\phi_l\right\rangle \left\langle\bar{\phi}_l|j\right\rangle\right],
\end{equation}
\begin{equation}
 \label{CoeffSijprime}
\mathcal{S}'_{ij}(\gamma)=-\sum_{l=2}^N\frac{\lambda_l}{[1-(1-\gamma)\lambda_l]^2}\left\langle i|\phi_l\right\rangle \left\langle\bar{\phi}_l|j\right\rangle .
\end{equation}
A numerical solution of Eq. (\ref{Opt_Cond1}) yields $\gamma^\star$: its evaluation involves sums of $N$ terms which depend on the eigenvalues and eigenvectors of the matrix $\mathbf{W}$ that defines the ergodic random walk without resetting and with stationary distribution $P_j^\infty(0)$, for $j=1,2,\ldots,N$. 

\subsection{Two general conditions for the first passage time fluctuations}
\label{Sec_OptimalReset_Analytical}
We have seen that Eq. (\ref{Opt_Cond1}) defines a general condition for the existence of a finite optimal resetting probability, and takes the form of different sums involving the eigenvalues and eigenvectors of the matrix $\mathbf{W}$. In this section, we further handle Eq. (\ref{Opt_Cond1}) to deduce a simpler condition, only in terms of the first two moments of the probability distribution of the first passage time. To this end, let us first consider the moments $\mathcal{R}_{ij}^{(1)}(i;\gamma)$ defined by Eq. (\ref{R1ij_spect}) in Appendix \ref{App_DeductionTij}. For the dynamics under resetting, we have
\begin{equation}
\mathcal{R}_{ij}^{(1)}(i;\gamma)=\sum_{l=2}^N\frac{(1-\gamma)\lambda_l}{(1-(1-\gamma)\lambda_l)^2}\left\langle i|\psi_l(i;\gamma)\right\rangle \left\langle\bar{\psi}_l(i;\gamma)|j\right\rangle,
\end{equation}
where $\left|\psi_l(i;\gamma)\right\rangle$ and $\left\langle\bar{\psi}_l(i;\gamma)\right|$ denote the right and left eigenvectors of $\mathbf{\Pi}(r;\gamma)$ with $r=i$. However, from the results of Section \ref{RW_reset1}, for $l=2,\ldots,N$ we have
\begin{eqnarray}
\nonumber
\displaystyle
&\left\langle j|\psi_l(i;\gamma)\right\rangle \left\langle\bar{\psi}_l(i;\gamma)|j\right\rangle-\left\langle i|\psi_l(i;\gamma)\right\rangle \left\langle\bar{\psi}_l(i;\gamma)|j\right\rangle\\
&
\displaystyle
=
\left\langle j|\phi_l\right\rangle \left\langle\bar{\phi}_l|j\right\rangle-\left\langle i|\phi_l\right\rangle \left\langle\bar{\phi}_l|j\right\rangle, 
\end{eqnarray}
therefore
\begin{equation}
\mathcal{R}_{jj}^{(1)}(i;\gamma)-\mathcal{R}_{ij}^{(1)}(i;\gamma)=\sum_{l=2}^N\frac{(1-\gamma)\lambda_l\left[
\left\langle j|\phi_l\right\rangle \left\langle\bar{\phi}_l|j\right\rangle-\left\langle i|\phi_l\right\rangle \left\langle\bar{\phi}_l|j\right\rangle\right]}{(1-(1-\gamma)\lambda_l)^2}.
\end{equation}
Consequently, $\mathcal{C}'_{ij}(\gamma)$ in Eq. (\ref{CoeffCijprime}) can be rewritten as
\begin{eqnarray}
\nonumber
\mathcal{C}'_{ij}(\gamma)&=-\frac{1}{1-\gamma}\sum_{l=2}^N\frac{(1-\gamma)\lambda_l}{[1-(1-\gamma)\lambda_l]^2}\left[
\left\langle j|\phi_l\right\rangle \left\langle\bar{\phi}_l|j\right\rangle-\left\langle i|\phi_l\right\rangle \left\langle\bar{\phi}_l|j\right\rangle\right]\\
\label{CijprimeR1}
&=-\frac{1}{1-\gamma}\left[\mathcal{R}_{jj}^{(1)}(i;\gamma)-\mathcal{R}_{ij}^{(1)}(i;\gamma)\right].
\end{eqnarray}    
In addition, from Eq. (\ref{Opt_Cond1}), the optimal $\gamma^\star$ obeys the equality
\begin{equation}\label{Opt_Cond2}
\mathcal{C}'_{ij}(\gamma^\star)=\langle T_{ij}(\gamma^\star)\rangle\left[\gamma^\star\, \mathcal{S}^\prime_{ij}(\gamma^\star)+\mathcal{S}_{ij}(\gamma^\star)\right].
\end{equation}    
Combining  Eqs. (\ref{CijprimeR1}) and (\ref{Opt_Cond2}), we get
\begin{equation}\label{Cdev_R1}
\mathcal{R}_{jj}^{(1)}(i;\gamma^\star)-\mathcal{R}_{ij}^{(1)}(i;\gamma^\star)=-(1-\gamma)
\langle T_{ij}(\gamma^\star)\rangle\left[\gamma^\star\, \mathcal{S}^\prime_{ij}(\gamma^\star)+\mathcal{S}_{ij}(\gamma^\star)\right].
\end{equation}
Let us now consider the second moment $\langle T^2_{ij}(\gamma)\rangle$ of the first passage time distribution. For $i\neq j$, this quantity reads (see the Appendix \ref{App_DeductionTij} for a detailed deduction)
\begin{eqnarray}
\nonumber
\langle T^2_{ij}(\gamma)\rangle=\frac{\mathcal{R}_{jj}^{(0)}(i;\gamma)-\mathcal{R}_{ij}^{(0)}(i;\gamma)}{P_j^\infty(i;\gamma)}\\\qquad+2\,\frac{ \mathcal{R}_{jj}^{(1)}(i;\gamma)-\mathcal{R}_{ij}^{(1)}(i;\gamma)}{ P_j^\infty(i;\gamma)}+2\,\frac{ \mathcal{R}_{jj}^{(0)}(i;\gamma) \left[\mathcal{R}_{jj}^{(0)}(i;\gamma)-\mathcal{R}_{ij}^{(0)}(i;\gamma)\right]}{ P_j^\infty(i;\gamma)^2}.
\end{eqnarray}
Reorganizing this expression and using $\langle T_{ij}(\gamma)\rangle$ given by Eq. (\ref{Tij}) in the Appendix  \ref{App_DeductionTij}, we obtain
\begin{equation}\label{Tijsquare_opt1}
\langle T^2_{ij}(\gamma)\rangle =\langle T_{ij}(\gamma)\rangle+2\,\frac{\mathcal{R}_{jj}^{(1)}(i;\gamma)-\mathcal{R}_{ij}^{(1)}(i;\gamma)}{ P_j^\infty(i;\gamma)}+2 \frac{\mathcal{R}_{jj}^{(0)}(i;\gamma)}{P_j^\infty(i;\gamma)} \langle T_{ij}(\gamma)\rangle.
\end{equation}
Using Eq. (\ref{Cdev_R1}), the above expression becomes
\begin{eqnarray}\nonumber
\langle T^2_{ij}(\gamma^\star)\rangle&=\langle T_{ij}(\gamma^\star)\rangle-2(1-\gamma^\star)
\,\frac{\gamma^\star\, \mathcal{S}^\prime_{ij}(\gamma^\star)+\mathcal{S}_{ij}(\gamma^\star)}{ P_j^\infty(i;\gamma^\star)}\langle T_{ij}(\gamma^\star)\rangle\\ &+2 \frac{\mathcal{R}_{jj}^{(0)}(i;\gamma^\star)}{P_j^\infty(i;\gamma^\star)} \langle T_{ij}(\gamma^\star)\rangle .
\end{eqnarray}
From the definition of $\mathcal{S}_{ij}(\gamma)$ in Eq. (\ref{CoeffSij}) and its derivative in Eq. (\ref{CoeffSijprime}), we get (for $i\neq j$)
\begin{eqnarray}\nonumber
&(1-\gamma)
\left[\gamma\, \mathcal{S}^\prime_{ij}(\gamma)+\mathcal{S}_{ij}(\gamma)\right]\\
\nonumber
&=(1-\gamma)\sum_{l=2}^N\left\{-\frac{\gamma\lambda_l}{[1-(1-\gamma)\lambda_l]^2}+\frac{1}{1-(1-\gamma)\lambda_l}
\right\}\left\langle i|\phi_l\right\rangle \left\langle\bar{\phi}_l|j\right\rangle\\ \label{condition_Sprime}
&=(1-\gamma)\sum_{l=2}^N\frac{1-\lambda_l}{[1-(1-\gamma)\lambda_l]^2}
\left\langle i|\phi_l\right\rangle \left\langle\bar{\phi}_l|j\right\rangle .
\end{eqnarray}
From the relations between the eigenvectors of $\mathbf{W}$ and $\mathbf{\Pi}(i,\gamma)$ (for $l=2,\ldots,N$),
\begin{eqnarray}\nonumber
\left\langle i|\psi_l(i;\gamma)\right\rangle \left\langle\bar{\psi}_l(i;\gamma)|j\right\rangle&=
\left[\left\langle i|\phi_l\right\rangle \left\langle\bar{\phi}_l|j\right\rangle-\gamma\frac{\left\langle i|\phi_l\right\rangle \left\langle\bar{\phi}_l|j\right\rangle}{1-(1-\gamma)\lambda_l} \right]\\
&=
\frac{(1-\gamma)(1-\lambda_l)}{1-(1-\gamma)\lambda_l}
\left\langle i|\phi_l\right\rangle \left\langle\bar{\phi}_l|j\right\rangle\qquad \mathrm{for}\, i\neq j.
\end{eqnarray}
Therefore, Eq. (\ref{condition_Sprime}) can be recast as
\begin{eqnarray}
\nonumber
&(1-\gamma)
\left[\gamma\, \mathcal{S}^\prime_{ij}(\gamma)+\mathcal{S}_{ij}(\gamma)\right]\\
&=\sum_{l=2}^N\frac{1}{1-(1-\gamma)\lambda_l}\left\langle i|\psi_l(i;\gamma)\right\rangle \left\langle\bar{\psi}_l(i;\gamma)|j\right\rangle=\mathcal{R}_{ij}^{(0)}(i;\gamma).
\end{eqnarray}
Hence, the identity (\ref{Tijsquare_opt1}) at optimality takes the much more compact form,
\begin{eqnarray}\nonumber
\langle T^2_{ij}(\gamma^\star)\rangle&=\langle T_{ij}(\gamma^\star)\rangle+2\, \frac{\mathcal{R}_{jj}^{(0)}(i;\gamma^\star)-\mathcal{R}_{ij}^{(0)}(i;\gamma^\star)}{P_j^\infty(i;\gamma^\star)} \langle T_{ij}(\gamma^\star)\rangle\\
&=\langle T_{ij}(\gamma^\star)\rangle+2\langle T_{ij}(\gamma^\star)\rangle^2 \qquad i\neq j, \label{Tijsquare_opt2}
\end{eqnarray}
where Eq. (\ref{Tij}) of the Appendix has been used in the last step.
At this point, it is convenient to introduce the coefficient of variation of the first passage time distribution (or re-scaled standard deviation), defined as 
\begin{equation}
z_{ij}(\gamma)\equiv\frac{\sqrt{\langle T^2_{ij}(\gamma)\rangle-\langle T_{ij}(\gamma)\rangle^2}}{\langle T_{ij}(\gamma)\rangle}.
\end{equation}
From Eq. (\ref{Tijsquare_opt2}), at the optimal resetting probability, the following equality holds
\begin{equation}\label{Z_condition_optimal}
z^2_{ij}(\gamma^\star)=1+\frac{1}{\langle T_{ij}(\gamma^\star)\rangle} \qquad \mathrm{for} \qquad i\neq j.
\end{equation}
A few comments are in order. Firstly, the coefficient of variation is not unity at 
$\gamma^\star$, contrary to processes defined in continuous time \cite{reuveni2016optimal,chechkin2018random,PalFP_PRL_2017,crit3}, but above. As the criterion (\ref{Z_condition_optimal}) involves the minimal MFPT explicitly, it depends on the network and process under consideration: therefore, the fluctuations of the first passage time at optimality are less universal. Secondly, recall that $\langle T_{ij}(\gamma^\star)\rangle$ is an adimensional quantity here, which represents an average number of steps to reach the target. The duration of a step being always fixed to 1, the continuous limit is recovered by letting the number of steps tend to infinity ($i$ and $j$ are very far apart, or $\langle T_{ij}(\gamma^\star)\rangle\gg 1$), which gives back the original criterion $z_{ij}=1$. 
\\[2mm]
Having found a general condition at optimal restart, we can follow a similar approach to derive another condition, that will tell us when the introduction of a small resetting probability improves an arbitrary Markov search process (with $\gamma=0$). Therefore, we impose that
\begin{equation}\label{D_Tij0}
\frac{d}{d\gamma}\langle T_{ij}(\gamma)\rangle\Big|_{\gamma\to 0}<0,
\end{equation}
which also implies that a non-zero $\gamma^\star$ exists. The latter statement follows from the fact that $\gamma^\star< 1$, since a process with $\gamma=1$ trivially stays on the initial node at all times and can never reach the target $j$ (or $\langle T_{ij}(\gamma)\rangle=\infty$). If Eq. (\ref{D_Tij0}) holds and assuming continuity, then $\langle T_{ij}(\gamma)\rangle$ is non-monotonic, with at least one minimum.
Using the expression (\ref{MFPTCS}), the condition above becomes
\begin{equation}
\frac{d}{d\gamma}\langle T_{ij}(\gamma)\rangle\Big|_{\gamma\to 0}=\frac{\mathcal{C}'_{ij}(0)P_j^\infty(0)-\mathcal{C}_{ij}(0)\mathcal{S}_{ij}(0)}{[P_j^\infty(0)]^2}<0,
\end{equation}
or
\begin{equation}\label{Cijprime_condition0}
\mathcal{C}'_{ij}(0)<\langle T_{ij}(0)\rangle\mathcal{S}_{ij}(0).
\end{equation}
In addition, we can combine Eqs. (\ref{CijprimeR1}) and (\ref{Tijsquare_opt1}) to obtain
\begin{equation}\label{Cijprime_condition0b}
2\frac{\mathcal{C}'_{ij}(0)}{P_j^\infty(0)}=\langle T_{ij}(0)\rangle-
\langle T^2_{ij}(0)\rangle+2\frac{\mathcal{R}_{jj}^{(0)}(i;0)}{P_j^\infty(0)}\langle T_{ij}(0)\rangle.
\end{equation}
Hence, the inequality (\ref{Cijprime_condition0}) can be rewritten as
\begin{equation}\label{Cijprime_condition1}
\langle T_{ij}(0)\rangle-
\langle T^2_{ij}(0)\rangle+2\frac{\mathcal{R}_{jj}^{(0)}(i;0)}{P_j^\infty(0)}\langle T_{ij}(0)\rangle<2\langle T_{ij}(0)\rangle\frac{\mathcal{S}_{ij}(0)}{P_j^\infty(0)}.
\end{equation}
Using Eq. (\ref{Tij}) and the  identity $\mathcal{S}_{ij}(0)=\mathcal{R}_{ij}^{(0)}(i;0)$ obtained from comparing Eqs. (\ref{CoeffSij}) and (\ref{Rij0_general}) with $\gamma=0$, Eq. (\ref{Cijprime_condition1}) becomes
\begin{equation}\label{Cijprime_condition1b}
\langle T_{ij}(0)\rangle-
\langle T^2_{ij}(0)\rangle+2\langle T_{ij}(0)\rangle^2<0,
\end{equation}
which is easily expressed as
\begin{equation}\label{Z_condition_gamma0}
z^2_{ij}(0)>1+\frac{1}{\langle T_{ij}(0)\rangle}\qquad \mathrm{for}\,\, i\neq j.
\end{equation}
The result (\ref{Z_condition_gamma0}) establishes a condition for which a given process will see its MFPT reduced by the introduction of a small amount of resetting.  When this second criteria holds, it also implies the existence of a non-zero optimal resetting probability $\gamma^\star$. We have thus deduced two related criteria that characterize the effect of resetting on Markov processes in discrete time. The first one is given by Eq. (\ref{Z_condition_optimal}) and quantifies the fluctuations at the optimal resetting probability, when it exists. These results are general and can be applied to any ergodic Markov process whose evolution is given by a master equation.
\\[2mm]
Notice that a resetting step always lasts one unit of time in the master equation (the same duration as a random walk step): as in similar models considered in the literature \cite{Kusmierz2014PRL}, resetting is thus not instantaneous here. The inequality (\ref{Z_condition_gamma0}) thus differs from another criterion recently derived for discrete-time processes by using a renewal method which implicitly assumed resetting events to have zero duration \cite{bonomo2021first}. In such a situation, the walker immediately jumps from its current position to the resetting point, meaning that it occupies two positions at the same discrete time $t$. It is therefore possible that the process ends when the steps for first passage and resetting coincide. The criterion for the underlying process ($\gamma=0$) in this case reads $z^2_{ij}(0)>1-1/\langle T_{ij}(0)\rangle$ \cite{bonomo2021first}. In a subsequent study \cite{bonomo2021polya}, the same authors have taken into consideration the duration of a resetting event in the renewal equations, obtaining the same criteria (\ref{Z_condition_gamma0}).
\section{Some examples}
\label{Sec_examples}
In this section, we explore the behaviour of the optimal resetting probability in three particular cases, namely, the standard random walk on rings, on Cayley trees, and L\'evy flights on rings.
\subsection{Standard random walks with resetting on rings}
\label{Normal_Rings}
Let us consider unbiased nearest-neighbour random walks (or standard random walks) on a network defined by its adjacency matrix $\mathbf{A}$, with elements $A_{ij}=A_{ji}=1$ if the nodes $i$ and $j$ are connected to each other and 0 otherwise. The quantity $k_i=\sum_{l=1}^NA_{il}$ is the degree of the node $i$. In this case, the transition probabilities are given by 
\begin{equation}
w_{i\to j}=\frac{A_{ij}}{k_i}.
\end{equation}
Therefore, the hops to the nodes connected to $i$ are equiprobable, whereas the transition between two nodes that are not directly connected by a link is not possible. 
\\[2mm]
In the following, we study the dynamics with stochastic resetting to the initial node on a ring of $N$ nodes. On this network, $k_i=2$ and the transition matrix $\mathbf{W}$ is a circulant matrix with well-known eigenvalues and eigenvectors \cite{FractionalBook2019,VanMieghem2011}. The eigenvalues are given by $\lambda_{l}=\cos\varphi_{l}$, with $\varphi_{l}\equiv\frac{2\pi}{N}(l-1)$, while the projections of the eigenvectors in the canonical base are $\langle j|\phi_{l}\rangle=\frac{1}{\sqrt{N}}e^{-\mathrm{i}\varphi_{l}(j-1)}$ and $\langle \bar{\phi}_{l}|j\rangle=\frac{1}{\sqrt{N}}e^{\mathrm{i}\varphi_{l}(j-1)}$, where $\mathrm{i}=\sqrt{-1}$. In addition, for this regular network, the stationary distribution is constant and given by $P_j^\infty(0)=1/N$.
\\[2mm]
Introducing this information on $\mathbf{W}$ into Eqs. (\ref{Pinfvectors_1})-(\ref{MFPT_resetSM}) allows us to obtain the stationary distribution
\begin{equation}
P_j^\infty(i,\gamma)=\frac{1}{N}+\frac{\gamma}{N}\sum_{l=2}^N \frac{\cos\left(\varphi_l\,d_{ij}\right)}{1-(1-\gamma)\cos(\varphi_l)}
\label{Pinf_ring}
\end{equation}
and the MFPT
\begin{equation}
\left\langle T_{ij}(\gamma)\right\rangle=\frac{1}{P_j^\infty(i,\gamma)}\left[\delta_{ij}+\frac{1}{N}
\sum_{l=2}^N\frac{1-\cos\left(\varphi_l\,d_{ij}\right)}{1-(1-\gamma)\cos(\varphi_l)}\right] \label{MFPT_ring}
\end{equation}
for the random walk with resetting to node $i$.
Here $d_{ij}$ is the length of the shortest path connecting the nodes $i$ and $j$, therefore $\cos\left[\frac{2\pi}{N}d_{ij}\right]=\cos\left[\frac{2\pi}{N}(i-j)\right]$ for any pair $\{i,j\}$ (in general $d_{ij}\neq |i-j|$).
Similar expressions can be deduced for the terms in Eq. (\ref{Opt_Cond1}), which is solved numerically for $\gamma^\star$.
\\[2mm]
\begin{figure*}[t!]
\centering
\includegraphics*[width=1.0\textwidth]{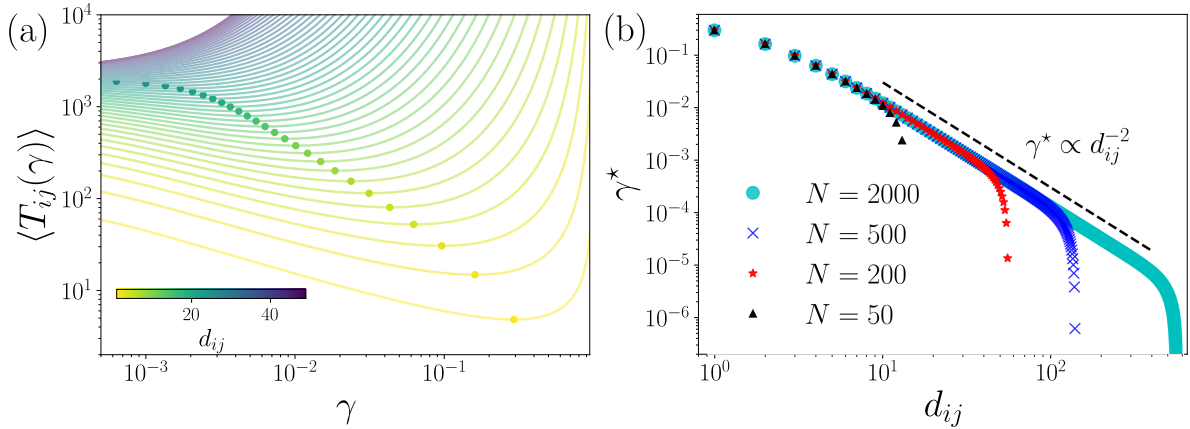}
\vspace{-5mm}
\caption{\label{Fig_1} Optimal resetting of standard random walks on rings. (a) MFPTs $\left\langle T_{ij}(\gamma)\right\rangle$ as a function of $\gamma$ for a ring with $N=100$. The curves correspond to different distances $d_{ij}$ (encoded in the colorbar) between the initial node $i$ where the walker is reset to and a target node $j$. The optimal resetting probabilities $\gamma^\star$ minimizing $\left\langle T_{ij}(\gamma)\right\rangle$ are represented with circles. (b) $\gamma^\star$ as a function of the distance $d_{ij}$ for rings of different sizes, obtained by solving numerically Eq. (\ref{Opt_Cond1}).
}
\end{figure*}
In Fig. \ref{Fig_1} we display the optimal resetting probability $\gamma^\star$ for rings of different sizes. In the panel \ref{Fig_1}a we present the values of $\left\langle T_{ij}(\gamma)\right\rangle$ for a ring with $N=100$ as a function of $\gamma$ and different distances between the initial node $i$ and the target $j$. The results are obtained using the analytical result (\ref{MFPT_ring}) and the resetting probabilities that minimize  $\left\langle T_{ij}(\gamma)\right\rangle$ in each curve are presented with circles. In the panel \ref{Fig_1}b we show the optimal $\gamma^\star$ obtained from Eq. (\ref{Opt_Cond1}) as a function of the distance $d_{ij}$ for rings of sizes $N=50,\,200,\,500,\,2000$.
\\[2mm]
In Fig. \ref{Fig_1}b we notice that for $N$ large and $d_{ij}\gg 1$, the optimal resetting probability satisfies the scaling law $\gamma^\star\propto  d_{ij}^{-2}$. This result is deduced analytically for an infinite ring $N\to \infty$ in \cite{ResetNetworks_PRE2020}. In the limit of less frequent resetting ($\gamma\ll 1$) and $d_{ij}\gg 1$, one recovers the expression of the continuous limit, $\langle T_{ij}(\gamma)\rangle \approx \frac{1}{\gamma}\left[e^{\sqrt{2\gamma} d_{ij}}-1\right]$ \cite{evans2011diffusion}, which is a non-monotonic function of $\gamma$. Solving $\partial \langle T_{ij}(\gamma)\rangle/\partial \gamma=0$ with this expression gives an optimal resetting probability $\gamma^\star\simeq 1.26982/d_{ij}^2$ for the search of a target located at a distance $d_{ij}\gg 1$ \cite{ResetNetworks_PRE2020}. This scaling relation can be explained with a simple argument. In a symmetric random walk (much like diffusion), the distance traveled by the walker is a result of fluctuations in the motion, namely, the distance is proportional to $\sqrt{s}$, where $s$ is the number of steps taken by the walker. Thus, in order to travel a distance $d_{ij}$, the walker should take on average $d_{ij}^2$ steps, before resetting occurs. Thus, $\gamma^\star\propto 1/d_{ij}^2$.
\subsection{Random walks on Cayley trees}
\label{Normal_Cayley}
\begin{figure*}[t!]
	\centering
	\includegraphics*[width=1.0\textwidth]{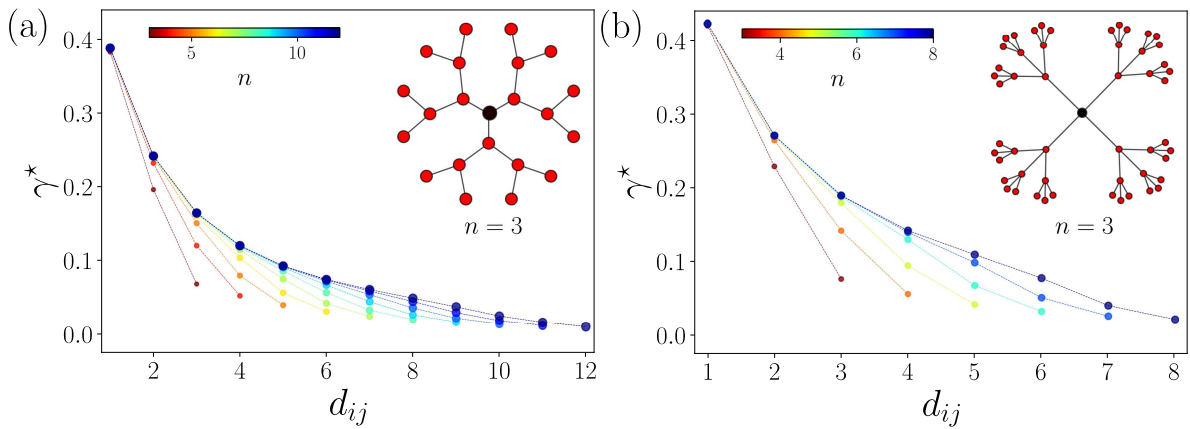}
	\vspace{-5mm}
	\caption{\label{Fig_2} Optimal searches by random walks with stochastic resetting to the central node on Cayley trees of $n$ shells and coordination number (a) $z=3$ and (b) $z=4$. We present the numerical value of the optimal resetting probability $\gamma^\star$  obtained by solving numerically Eq. (\ref{Opt_Cond1}). The values are shown as a function of the distance $d_{ij}$ between the initial/resetting node $i$ and a target node $j$ (continuous lines are used as a guide). The number of shells, $n$, is varied as encoded in the colorbar. We show in the insets the trees with $n=3$.
	}
\end{figure*}
We now consider a nearest-neighbour random walk on finite Cayley trees of coordination number $z$ and of $n$ shells (see two examples in Fig. \ref{Fig_2} for $n=3$ and $z=3$ and $4$). In this case, the nodes of the last shell have degree $1$, whereas the other nodes have degree $z$. Recall that the stationary distribution of the walker without resetting is given by $P_j^\infty(0)=k_j/\sum_{l=1}^N k_l$ (see \cite{NohRieger2004} for details).
\\[2mm]
We display $\gamma^\star$ for trees of varying number of shells $n$ and coordination number $z=3$ or $4$, see Figs. \ref{Fig_2}a and b, respectively. The starting/resetting position $i$ is the central node in all cases. In these examples, the eigenvalues and eigenvectors of the matrix $\mathbf{W}$ are obtained numerically and plugged into the sums of Eq. (\ref{Opt_Cond1}) to obtain $\gamma^\star$  numerically.
\\[2mm]
We notice that $\gamma^\star$ slowly decays with $d_{ij}$. At large $n$ and for $1\ll d_{ij}\ll n$, 
\begin{equation}\label{gammaopt}
\gamma^\star\simeq \frac{1}{d_{ij}}\left(\frac{z-2}{z}\right).
\end{equation}
See \cite{ResetNetworks_PRE2020} for a deduction of this result.
Hence the optimal resetting rate tends to 0 at large $d_{ij}$ differently than on $d$-dimensional lattices, where $\gamma^*\sim 1/d_{ij}^2$ as we have seen in the case of the ring. This is due to the fact that random walks on Cayley trees are effectively drifting away from their starting point \cite{cassi1989random} and thus travel a distance $d$ during a time of order $d$, instead of $d^2$.
\\[2mm]
\begin{figure*}[t!]
	\centering
	\includegraphics*[width=1.0\textwidth]{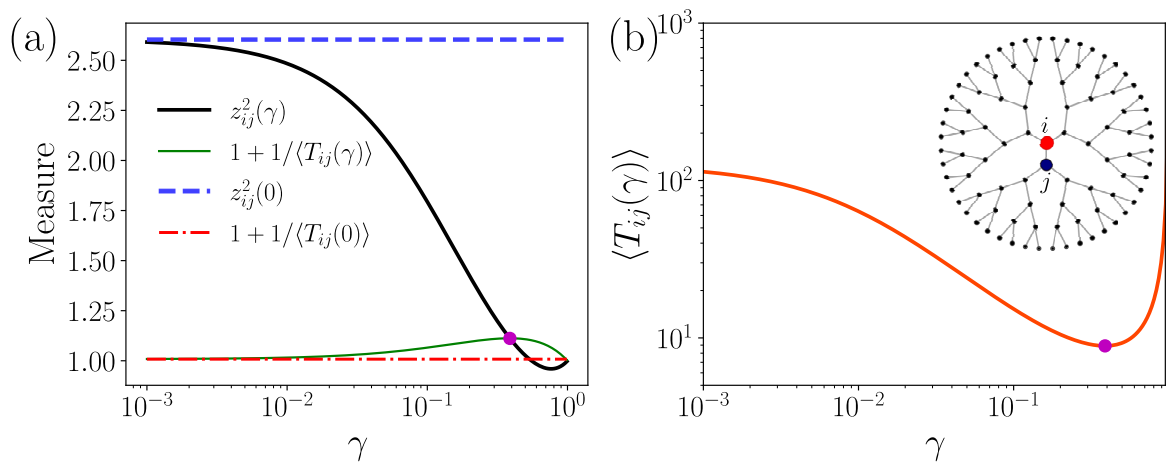}
	\vspace{-4mm}
	\caption{\label{Fig_3}  Optimal resetting on a Cayley tree with coordination number $z=3$ and $n=4$ shells, in a case where $1+1/\langle T_{ij}(0)\rangle<z^2_{ij}(0)$. (a) $z^2_{ij}(\gamma)$ and $1+1/\langle T_{ij}(\gamma)\rangle$ as a function of $\gamma$. The intersection of these curves is represented with a dot and occurs at $\gamma=\gamma^\star$. We also show the values of $1+1/\langle T_{ij}(0)\rangle$ and $z^2_{ij}(0)$, showing that the condition (\ref{Z_condition_gamma0}) is satisfied. (b) $\langle T_{ij}(\gamma)\rangle$ as a function of $\gamma$, where $\gamma^\star$ is represented by a dot and corresponds to the minimum. Inset: network with the initial/resetting node $i$ and the target node $j$ marked.
	}
\end{figure*}
\begin{figure*}[t!]
	\centering
	\includegraphics*[width=1.0\textwidth]{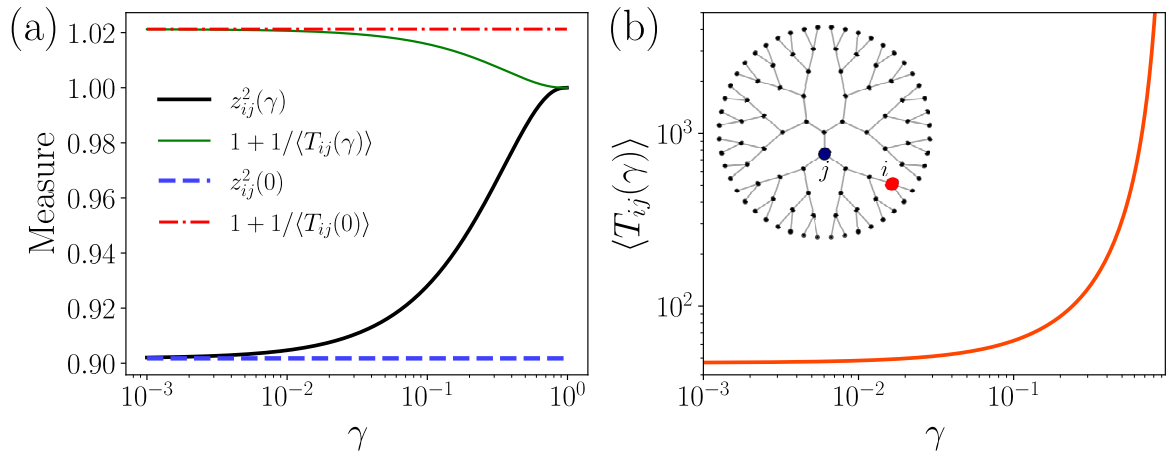}
	\vspace{-4mm}
	\caption{\label{Fig_4} Effect of resetting on the Cayley tree of Fig. \ref{Fig_3}, in a case where $1+1/\langle T_{ij}(0)\rangle>z^2_{ij}(0)$. (a) $z^2_{ij}(\gamma)$ and $1+1/\langle T_{ij}(\gamma)\rangle$ as a function of $\gamma$. In this case, the curves do not intersect for $\gamma<1$. (b) $\langle T_{ij}(\gamma)\rangle$ as a function of $\gamma$. Inset: network with the initial/resetting node $i$ and the target node $j$ marked.
	}
\end{figure*}
In order to illustrate the conditions (\ref{Z_condition_optimal}) and (\ref{Z_condition_gamma0}), let us consider a normal random walk on a Cayley tree with $z=3$ and $n=4$. In Figs. \ref{Fig_3} and \ref{Fig_4}, we analyze the dynamics with resetting for two particular pairs of starting and target nodes, depicted in the insets of Figs. \ref{Fig_3}b and \ref{Fig_4}b, respectively.  
\\[2mm]
Fig. \ref{Fig_3} corresponds to a case where the condition (\ref{Z_condition_gamma0}) is satisfied, as illustrated by the two horizontal dashed lines of the panel \ref{Fig_3}a. We see in Fig. \ref{Fig_3}b how $\langle T_{ij}(\gamma)\rangle$ actually decreases as $\gamma$ increases, until reaching a minimum at $\gamma^\star$. At this same value of $\gamma$, the condition (\ref{Z_condition_optimal}) is fulfilled, as illustrated by Fig. \ref{Fig_3}a, where the curve $1+1/\langle T_{ij}(\gamma)\rangle$ intersects $z^2_{ij}(\gamma)$. 
\\[2mm]
In contrast, in Fig. \ref{Fig_4}, $i$ and $j$ are such that $1+1/\langle T_{ij}(0)\rangle>z^2_{ij}(0)$ (as shown by the two horizontal dashed lines of Fig. \ref{Fig_4}a). Consequently, the condition (\ref{Z_condition_gamma0}) is not satisfied and the MFPT keeps increasing with $\gamma$, with $\langle T_{ij}(\gamma)\rangle>\langle T_{ij}(0)\rangle$ for any $\gamma>0$, see Fig. \ref{Fig_4}b. We check in Fig. \ref{Fig_4}a that the curves $1+1/\langle T_{ij}(\gamma)\rangle$ and $z^2_{ij}(\gamma)$ do not intersect, except at the singular point $\gamma=1$.

\subsection{L\'evy flights with resetting on rings}
L\'evy flights on an arbitrary network can be generated using real powers of the Laplacian matrix $\mathbf{L}$, whose elements are $L_{ij}=\delta_{ij} k_i-A_{ij}$. The transition probabilities are \cite{RiascosMateosFD2014}
\begin{equation}\label{wijfrac}
w_{i\to j}(\alpha)=\delta_{ij}-\frac{(\mathbf{L}^\alpha)_{ij}}{(\mathbf{L}^\alpha)_{ii}}\qquad 0<\alpha< 1.
\end{equation}
In this definition $\mathbf{L}^\alpha$ is the fractional graph Laplacian introduced in Ref. \cite{RiascosMateosFD2014}. 
\begin{figure*}[t!]
\centering
\includegraphics*[width=0.65\textwidth]{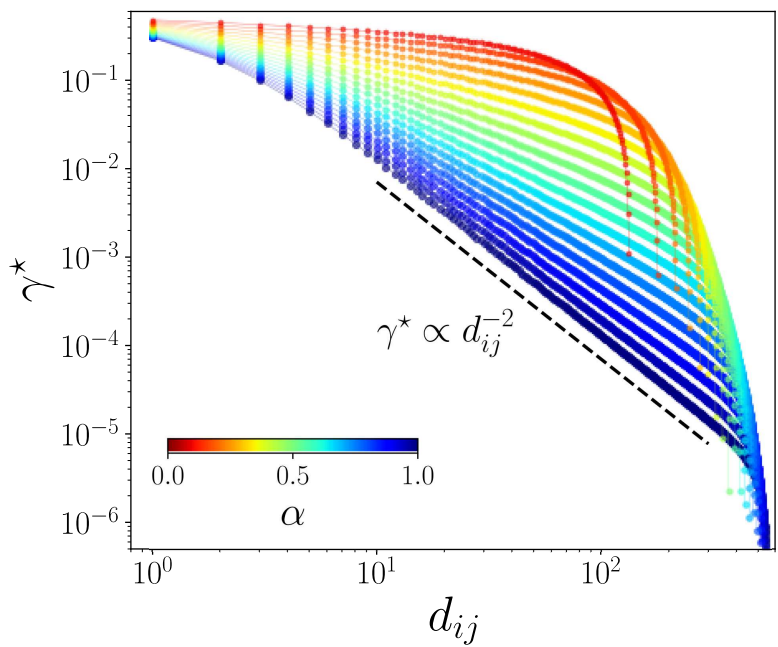}
\vspace{2mm}
\caption{\label{Fig_5}  Optimal resetting of L\'evy flights on a ring of $N=2000$ nodes. We present numerical results for $\gamma^\star$ (that minimizes $\left\langle T_{ij}(\gamma)\right\rangle$) as a function of the distance $d_{ij}$ between the source $i$ and the target node $j$ for processes defined by Eq. (\ref{wijfrac}), the results are obtained by solving numerically Eq. (\ref{Opt_Cond1}). The value of $\alpha$ is varied in the interval $[0.1,1]$, see the color code. In the case $\alpha=1$, the scaling behaviour of the local random walker, displayed in Fig. \ref{Fig_1}b, is recovered.
}
\end{figure*}
\begin{figure*}[t!]
	\centering
	\includegraphics*[width=1.0\textwidth]{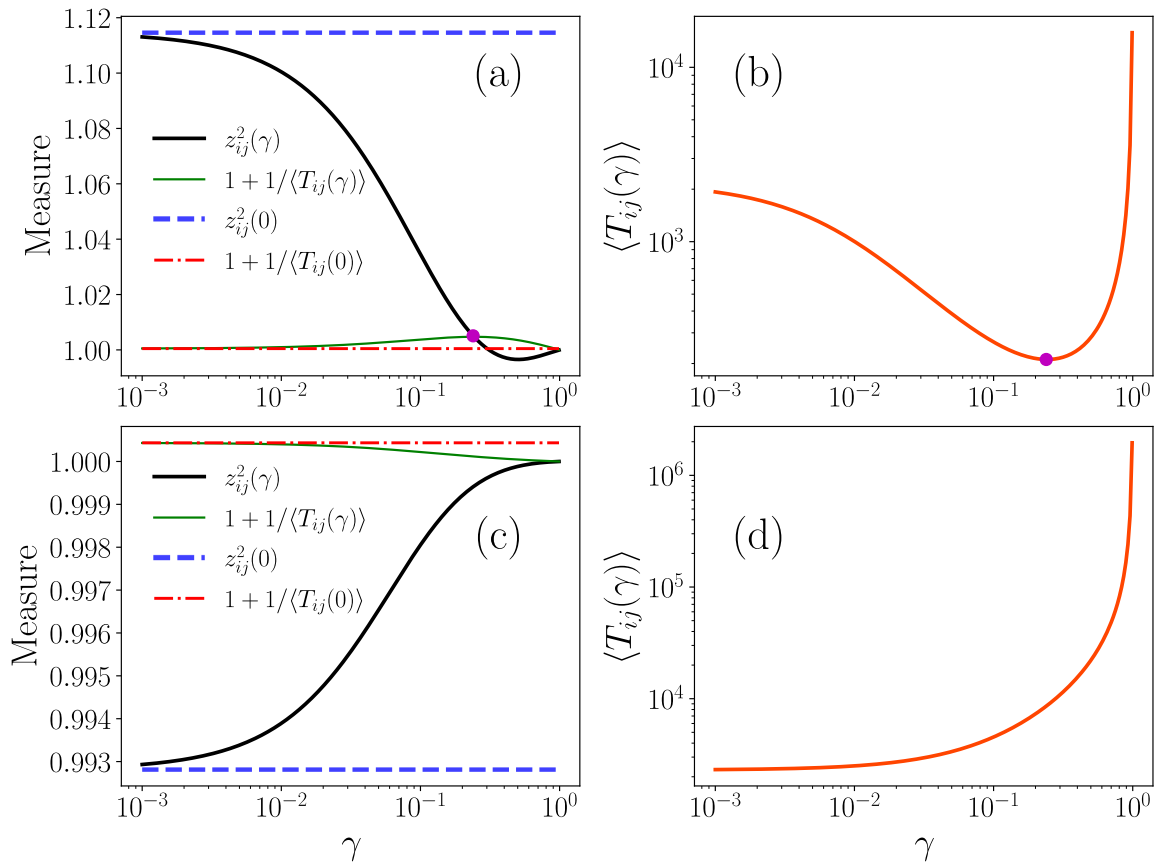}
	\vspace{-5mm}
	\caption{\label{Fig_6}  Conditions for optimal reset of L\'evy flights with $\alpha=0.2$ on a ring with $N=2000$ nodes. Two distances $d_{ij}$ between the initial and target nodes ($i$ and $j$, respectively) are considered: $d_{ij}=10$  (Panels (a)-(b)) and $d_{ij}=500$ in (Panels (c)-(d)). The panels (a) and (c) display $z^2_{ij}(\gamma)$ and $1+1/\langle T_{ij}(\gamma)\rangle$ as a function of $\gamma$. The values for $\gamma=0$ are marked with horizontal lines. The panels (b) and (d) show $\langle T_{ij}(\gamma)\rangle$ as a function of $\gamma$; in (a), the optimal $\gamma^\star$ is represented by a dot.
	}
\end{figure*}
In the limit $\alpha\to 1$, the simple random walk with transitions to nearest-neighbor nodes is recovered. In the case of regular networks with degree $k$, the elements of the fractional Laplacian take the form \cite{ReviewJCN_2021,RiascosMateosFD2015} 
\begin{equation}\label{Frac_Lap_regular}
(\mathbf{L}^{\alpha})_{ij}
=\sum_{m=0}^\infty {\alpha \choose m}(-1)^m k^{\alpha-m}(\mathbf{A}^m)_{ij},
\end{equation}
where ${x \choose y}\equiv\frac{\Gamma(x+1)}{\Gamma(y+1)\Gamma(x-y+1)}$ with $\Gamma(x)$ the Gamma function, and $(\mathbf{A}^m)_{ij}$  
is the number of all the possible paths of length $m$ connecting the nodes $i$, $j$. The relation (\ref{Frac_Lap_regular}) shows that the fractional Laplacian in regular networks includes information of all the possible paths connecting two nodes of the network. In the case of rings (regular networks with degree $k=2$), the transition probabilities in Eq. (\ref{wijfrac}) with $0<\alpha<1$ define a non-local dynamics with 
$w_{i\to j}(\alpha)\sim d_{ij}^{-(1+2\alpha)}$, $0<\alpha< 1$. Here, $d_{ij}$ is the length of the shortest path between $i$ and $j$, and $d_{ij}\gg 1$. In this manner, the jumps generated with the fractional Laplacian have the properties of a L\'evy flight with index $\mu=2\alpha$ (see Refs. \cite{ReviewJCN_2021,RiascosMateos2012,FractionalBook2019,RiascosMateosFD2015} for a detailed discussion on L\'evy flights and fractional transport on networks). 
\\[2mm]
In the case of finite rings with $N$ nodes, the Laplacian $\mathbf{L}$, the fractional Laplacian $\mathbf{L}^\alpha$ and the transition matrix $\mathbf{W}$ with elements given by Eq. (\ref{wijfrac}), are also circulant matrices \cite{FractionalBook2019,RiascosMateosFD2015}. Therefore, the left and right eigenvectors are the same as the ones exposed in Section \ref{Normal_Rings}. The eigenvalues $\{\lambda_l(\alpha)\}_{l=1} ^N$ of the transition matrix defined in Eq. (\ref{wijfrac}) are different and given by \cite{RiascosMateosFD2015,RiascosMichelitsch2017_gL}
\begin{equation}\label{lambdaGLring}
\lambda_l(\alpha)=1-\frac{1}{k^{(\alpha)}}\, \left(2-2\cos\varphi_l\right)^\alpha
\end{equation}
where $\varphi_l=\frac{2\pi}{N}(l-1)$ and $k^{(\alpha)}$ is the so-called {\it fractional degree}, defined as $k^{(\alpha)}=\frac{1}{N}\sum_{l=1}^N \left(2-2\cos\varphi_l\right)^\alpha$ \cite{RiascosMateosFD2015,RiascosMichelitsch2017_gL}.
\\[2mm]
The stationary distribution $P_j^\infty(i,\gamma)$ and the MFPT $\left\langle T_{ij}(\gamma)\right\rangle$ for L\'evy flights with resetting on rings are thus obtained simply by substituting $\lambda_l(0)=\cos(\varphi_l)$ by $\lambda_l(\alpha)$ in the expressions of the Section \ref{Normal_Rings}. They read (see \cite{MultipleResetPRE_2021} for more details)
\begin{equation}
P_j^\infty(i,\gamma)=\frac{1}{N}+\frac{\gamma}{N}\sum_{l=2}^N \frac{\cos\left(\varphi_l\,d_{ij}\right)}{1-(1-\gamma)\lambda_l(\alpha)},
\label{Pinf_ring_levy}
\end{equation}
and
\begin{equation}
\left\langle T_{ij}(\gamma)\right\rangle=\frac{1}{P_j^\infty(i,\gamma)}\left[\delta_{ij}+\frac{1}{N}
\sum_{l=2}^N\frac{1-\cos\left(\varphi_l\,d_{ij}\right)}{1-(1-\gamma)\lambda_l(\alpha)}\right]. \label{MFPT_ring_levy}
\end{equation}
The sums that are required to solve numerically for $\gamma^\star$ in Eq. (\ref{Opt_Cond1}) are evaluated using the eigenvalues $\lambda_l(\alpha)$ and the eigenvectors of circulant matrices. In Fig. \ref{Fig_5}, we display the optimal resetting probability $\gamma^\star$ as a function of the distance $d_{ij}$ on a ring of $N=2000$ nodes, for L\'evy flights generated by Eq. (\ref{wijfrac}) with different values of $\alpha$. The results show how the value of $\alpha$ that defines the long-range dynamics affects the decay of $\gamma^\star$ with $d_{ij}$. In the limit $\alpha\to 1$, one recovers the scaling of the local random walk discussed in Fig. \ref{Fig_1}b with $\gamma^\star \propto d_{ij}^{-2}$.
\\[2mm]
In Fig. \ref{Fig_6} we test the two conditions for optimal resetting derived in Sec. \ref{Sec_OptimalReset_Analytical} when the random walker follows a L\'evy flight defined by Eq. (\ref{wijfrac}). We choose $\alpha=0.2$ and consider two values of the distance $d_{ij}$ between the initial and the target nodes, on a ring with $N=2000$.  Figs. \ref{Fig_6}(a)-(b) correspond to $d_{ij}=10$. The panel \ref{Fig_6}(a) represents the different quantities involved in the two criteria: in this case  $1+1/\langle T_{ij}(0)\rangle<z^2_{ij}(0)$, hence a finite resetting probability optimizes the MFPT. The calculation of $z^2_{ij}(\gamma)$ and $1+1/\langle T_{ij}(\gamma)\rangle$ as a function of $\gamma$ shows how the two curves intersect at $\gamma^\star=0.2425$, where the condition (\ref{Z_condition_optimal}) is met. In Fig. \ref{Fig_6}(b), $\langle T_{ij}(\gamma)\rangle$ has a non-monotonic behaviour with a minimum at $\gamma^\star$. The same analysis is repeated in Figs. \ref{Fig_6}(c)-(d) for $d_{ij}=500$, a case where $1+1/\langle T_{ij}(0)\rangle>z^2_{ij}(0)$ (panel \ref{Fig_6}(c)). Consequently, resetting does not improve the L\'evy strategy for reaching the target node. This is confirmed in in Fig. \ref{Fig_6}(d), where $\langle T_{ij}(\gamma)\rangle$ is monotonic and increases with $\gamma$.
\\[2mm]
The results in this section illustrate the generality of the approach exposed in Sec. \ref{Sec_OptimalReset_Analytical}, allowing us to analyze optimal resetting in different random walk scenarios, either local, as in the examples of Secs. \ref{Normal_Rings} and \ref{Normal_Cayley}, or non-local as in the case of L\'evy flights.

\section{Conclusions}\label{sec:conclusions}
We have deduced some conditions for the existence of optimal resetting in discrete-time random walks on arbitrary networks, by using the eigenvalues and eigenvectors of the transition matrix that defines the dynamics without resetting. First, we derived a general condition fulfilled by the optimal resetting probability $\gamma^\star$, obtained from minimising the MFPT at a target node $j$ from an initial/resetting node $i$, or $\frac{d}{d\gamma} \left\langle T_{ij}(\gamma)\right\rangle\Big|_{\gamma=\gamma^\star}=0$. This condition allowed us to obtain numerically the values of $\gamma^\star$ in a number of cases. Secondly, we deduced a general expression for $\left\langle T^2_{ij}(\gamma)\right\rangle$. This quantity allowed us to obtain a general identity, see Eq. (\ref{Z_condition_optimal}), which relates the coefficient of variation of the first passage time and its mean at optimality. Another condition was deduced from imposing $\frac{d}{d\gamma} \left\langle T_{ij}(\gamma)\right\rangle\Big|_{\gamma=0}<0$, which represents the situation where resetting reduces the MFPT of a given process. The result is given by the inequality (\ref{Z_condition_gamma0}), which depends on the mean and variance of the first passage time of the process without resetting. We illustrated the results with the study of the optimal transport of random walks on rings and Cayley trees. We also discussed L\'evy flights with restart on rings.
The formalism exposed here is general and can be applied to any discrete-time ergodic Markovian process.
\\[2mm]
The question of optimal resetting on networks when there are several resetting nodes has not been addressed yet, despite recent studies on the subject \cite{MultipleResetPRE_2021}. More complicated situations may arise in those cases, such as the presence of various minima and metastability effects, as suggested by recent results in continuous spaces \cite{ciliberto,mercado}.

\section{Appendix: MFPTs for ergodic random walks}
\label{App_DeductionTij}
We present the deduction of the first and second moments of the first passage time distribution of random walks on networks defined by a transition probability matrix $\mathbf{\Pi}$ with elements $\pi_{i\to j}$ and leading to a stationary distribution $P_j^\infty$, $j=1,2,\ldots,N$. The results are general and apply to any ergodic Markovian process \cite{Hughes,NohRieger2004}. The discrete-time master equation $P_{ij}(t+1) = \sum_{m=1}^N  P_{im} (t) \pi_{m\rightarrow j}$ describes the evolution of the occupation probability $P_{ij}(t)$ of a random walker at $j$ at time $t$, given a starting node $i$ \cite{Hughes}.
The occupation probability can be also expressed as \cite{Hughes,NohRieger2004}
\begin{equation}\label{EquF}
P_{ij}(t) = \delta_{t0} \delta_{ij} + \sum_{t'=0}^t   P_{jj}(t-t')  F_{ij}(t') \ ,
\end{equation}
where $F_{ij}(t^\prime)$ is the probability to reach the node $j$ for the first time after exactly $t^\prime$ steps, starting from $i$. By definition $F_{ij}(0)=0$, and $P_{jj}(t-t^\prime)$ is the probability to be located at the position $j$ again after $t-t^\prime$ steps. The first term in the right-hand side of Eq. (\ref{EquF}) enforces the initial condition.
\\[2mm]
Introducing the discrete Laplace transform $\tilde{f}(s) \equiv\sum_{t=0}^\infty e^{-st} f(t)$ in Eq. (\ref{EquF}), we have
\begin{equation}\label{LaplTransF}
\widetilde{F}_{ij} (s) = (\widetilde{P}_{ij}(s) - \delta_{ij}) /
\widetilde{P}_{jj} (s) \ .
\end{equation}
In terms of $F_{ij}(t)$, the mean first passage time $\langle T_{ij}\rangle$ is given by \cite{Hughes}
\begin{equation}
\langle T_{ij}\rangle \equiv \sum_{t=0}^{\infty} t F_{ij} (t) = -\widetilde{F}'_{ij}(0).
\end{equation}
Likewise, the second moment $\langle T^2_{ij}\rangle$ of the probability distribution of the first passage time is given by
\begin{equation}
\langle T^2_{ij}\rangle \equiv \sum_{t=0}^{\infty} t^2 F_{ij} (t) =\widetilde{F}''_{ij}(0).
\end{equation}
Using the moments $\mathcal{R}^{(n)}_{ij}$ of $P_{ij}(t)$, defined as
\begin{equation}
\mathcal{R}^{(n)}_{ij}\equiv \sum_{t=0}^{\infty} t^n ~ \{P_{ij}(t)-P_j^\infty\},
\end{equation}
one can write the following expansion of $\widetilde{P}_{ij}(s)$,
\begin{equation}
\widetilde{P}_{ij}(s) =P_j^\infty\frac{1}{1-e^{-s}}
+ \sum_{n=0}^\infty (-1)^n \mathcal{R}^{(n)}_{ij} \frac{s^n}{n!} \ .
\end{equation}
Introducing this result into Eq. (\ref{LaplTransF}), we get
\begin{eqnarray*}\nonumber
&\hat{F}_{ij}(s)=1-s \frac{\mathcal{R}_{jj}^{(0)}-\mathcal{R}_{ij}^{(0)}+\delta _{ij}}{P_j^\infty}\\&\hspace{-2.6cm}+\frac{s^2}{2} \frac{(P_j^\infty+2 \mathcal{R}_{jj}^{(0)}) \delta _{ij}+P_j^\infty (\mathcal{R}_{jj}^{(0)}-\mathcal{R}_{ij}^{(0)}+2 (\mathcal{R}_{jj}^{(1)}-\mathcal{R}_{ij}^{(1)}))+2 \mathcal{R}_{jj}^{(0)} (\mathcal{R}_{jj}^{(0)}-\mathcal{R}_{ij}^{(0)})}{ (P_j^\infty)^2}+O(s^3).
\end{eqnarray*}
Taking the first two derivatives of this expression with respect to $s$, one obtains
\begin{equation}\label{Tij}
\langle T_{ij} \rangle =\frac{1}{P_j^\infty}\left[\delta_{ij}+\mathcal{R}^{(0)}_{jj}-\mathcal{R}^{(0)}_{ij}\right] ,
\end{equation}
and
\begin{eqnarray}\nonumber
\langle T^2_{ij} \rangle =\frac{(P_j^\infty+2 \mathcal{R}_{jj}^{(0)}) \delta _{ij}}{ (P_j^\infty)^2}&+\frac{\mathcal{R}_{jj}^{(0)}-\mathcal{R}_{ij}^{(0)}}{ P_j^\infty}
\\&+2\frac{\mathcal{R}_{jj}^{(1)}-\mathcal{R}_{ij}^{(1)}}{ P_j^\infty}+2\frac{ \mathcal{R}_{jj}^{(0)} (\mathcal{R}_{jj}^{(0)}-\mathcal{R}_{ij}^{(0)})}{ (P_j^\infty)^2},
\label{T2ij}
\end{eqnarray}
which are the expressions used in Section \ref{Sec_OptimalReset_Analytical}.
\\[2mm]
We may further use the spectral form of $\mathbf{\Pi}$ to rewrite the above expressions for $\langle T_{ij} \rangle$ and $\langle T^2_{ij} \rangle$ in a different way. The eigenvalues of the matrix $\mathbf{\Pi}$ are $\zeta_l$ (where $\zeta_1=1$), and its right and left eigenvectors are $\left|\psi_l\right\rangle$ and $\left\langle\bar{\psi}_l\right|$, respectively, for $l=1,2,\ldots,N$. 
Representing the transition matrix as $\mathbf{\Pi}=\sum_{l=1}^N \zeta_l|\psi_l\rangle\langle \bar{\psi}_l|$ and the occupation probability as $P_{ij}(t)=\langle i|\mathbf{\Pi}^t|j\rangle$, the moments $\mathcal{R}_{ij}^{(0)}$ and $\mathcal{R}_{ij}^{(1)}$ take the form
\begin{eqnarray}\nonumber
\mathcal{R}_{ij}^{(0)}&=\sum_{t=0}^\infty (P_{ij}(t)-P_j^{\infty})=\sum_{t=0}^\infty \sum_{m=2}^N \zeta_m^t \left\langle i|\psi_m\right\rangle \left\langle\bar{\psi}_m|j\right\rangle\\
&
=\sum_{m=2}^N\frac{1}{1-\zeta_m}\left\langle i|\psi_m\right\rangle \left\langle\bar{\psi}_m|j\right\rangle,
\label{Rij0_general}
\end{eqnarray}
and
\begin{eqnarray}\nonumber
\mathcal{R}_{ij}^{(1)}&=\sum_{t=0}^\infty t(P_{ij}(t)-P_j^{\infty})=\sum_{t=0}^\infty \sum_{m=2}^N t\zeta_m^t \left\langle i|\psi_m\right\rangle \left\langle\bar{\psi}_m|j\right\rangle\\
&
=\sum_{m=2}^N\frac{\zeta_m}{(1-\zeta_m)^2}\left\langle i|\psi_m\right\rangle \left\langle\bar{\psi}_m|j\right\rangle.
\label{R1ij_spect}
\end{eqnarray}
The final expressions for $\langle T_{ij} \rangle$ and $\langle T^2_{ij} \rangle$ for a random walker defined by the transition matrix $\mathbf{\Pi}$ are thus obtained by substituting Eqs. (\ref{Rij0_general}) and (\ref{R1ij_spect}) into the relations (\ref{Tij}) and (\ref{T2ij}).
\section*{Acknowledgments}
APR and DB acknowledge support from Ciencia de Frontera 2019 (CONACYT), project ``Sistemas complejos estoc\'asticos: Agentes m\'oviles, difusi\'on de part\'iculas, y din\'amica de espines'' (Grant No. 10872).

\section*{References}

\providecommand{\newblock}{}

\end{document}